\documentclass[12pt]{iopart}
\usepackage{amssymb}
\usepackage{graphicx}
\newcommand{\intL}{\int_0^L}
\newcommand{\vF}{v_\mathrm{F}^{{}}}
\newcommand{\nm}{\mathrm{nm}}
\newcommand{\eqref}[1]{equation~(\ref{#1})}

\begin{document}

\title{Electron-vibron coupling in suspended nanotubes}
\author{Karsten Flensberg}
\address{Nano-Science Center, Niels Bohr Institute,\\
Universitetsparken 5, 2100 Copenhagen, Denmark}
\ead{flensberg@fys.ku.dk}
\date{30 September 2005}

\begin{abstract}
We consider the electron-vibron coupling in suspended  nanotube
quantum dots. Modelling the tube as an elastic medium, we study the
possible coupling mechanism for exciting the stretching mode in a
single-electron-transistor setup. Both the forces due to the
longitudinal and the transverse fields are included. The effect of
the longitudinal field is found to be too small to be seen in
experiment. In contrast, the transverse field which couple to the
stretching mode via the bending of the tube can in some cases give
sizeable Franck-Condon factors. However, the length dependence is
not compatible with recent experiments [Sapmaz et al.
cond-mat/0508270].
\end{abstract}

\section{Introduction}

Suspended nanotubes form a interesting and promising system for
various nanoelectromechanical device setups and has been studied
both experimentally\cite{nygard01,leroy05,sapm05} and
theoretically\cite{kina03,jons04,jons05,sapm05,ustu05,izum05}.
Because of their large aspect ratio, nanotubes can be modeled as
simple one-dimensional strings using classical elasticity
theory\cite{suzu02}. Here we study the electromechanical coupling
when suspended nanotubes are put in a single-electron-transistor
(SET) setup.

When nanotubes are contacted by electrodes they form in most cases
contacts with a large resistance, which results in Coulomb blockade
behavior. This type of single-electron-tunnelling devices have also
been fabricated with the nanotubes suspended between the two
electrodes\cite{nygard01,leroy05,sapm05}. For these devices the
interesting possibility occurs that a coupling between the
electronic degree of freedom and the vibration might show up in the
current. Such a coupling has indeed been observed in several
experiments. The first example LeRoy et al.\cite{leroy05} observed
phonon sidebands with both absorbtion and emission peaks, which were
taken as evidence for the radial breathing mode being strongly
excited and thus behaving essentially as a classical external
time-dependent potential.

In the quantum regime the electron-vibron coupling leads to steps in
the IV characteristic, similar to the well-known Franck-Condon
principle. This has been seen in a number of single-molecule
devices\cite{park00,pasu05} and is well understood in terms of rate
equations\cite{boes01,mcca03,brai03flen,koch05}. Recently, similar
physics was observed in suspended nanotubes\cite{sapm05} where the
vibrational energy suggested that the sidebands were due to coupling
to a longitudinal stretching mode. However, the coupling mechanism
was unclear. It was suggested in Ref. \cite{sapm05}, that the
electric field parallel to the tube coupled to the nuclear motion.
Here we will argue that due to screening in the tube the
longitudinal electric field is too weak to excite the longitudinal
mode, and instead point to the non-linear coupling between the
traverse and longitudinal mode as the possible coupling mechanism.

The paper is organized as follows. First, we discuss in
section~\ref{sec:electrostat} the electrostatics of the charged
suspended nanotube, followed by an account in
section~\ref{sec:elastic} of the elastic properties of the hanging
tube. In section~\ref{sec:FC}, the modes of the tube are quantized
and the Franck-Condon overlap factors are calculated. Finally,
conclusions are given in section~\ref{sec:disc}.
\begin{figure}
  \centerline{\includegraphics[width=0.5\textwidth]{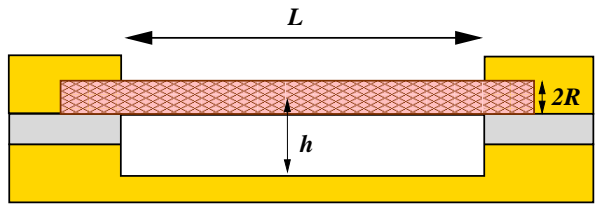}}
  \caption{Illustration of suspended nanotube device.}
  \label{fig:cnt}
\end{figure}

\section{Electrostatics of charge nanotube}
\label{sec:electrostat}

We now discuss the electrostatics of the charge suspended nanotube.
First we consider the longitudinal electric field, and then the
radial field.

\subsection{Electric field parallel to the nanotube}
\label{sec:Epar}

For a metallic tube there would, of course, be no electric field in
the longitudinal direction. However, the tube has a finite screening
length due to the kinetic energy. We have analyzed this situation
using the following density functional for the total energy of a
nanotube with linear dispersion
\begin{equation}\label{Frho}
    F[\rho]=
    \frac{\hbar}{2}\intL dx\,\vF(x)[\rho(x)]^2+\frac12\intL dx\intL
    dx'\,\rho(x)V(x,x')\rho(x'),
\end{equation}
where $v_F$ is the Fermi velocity and $V(x,x')$ is the effective 1D
potential for a cylindric conductor. Details about the interaction
and the solution are given in Appendix A. One should include
screening due to both gate and source-drain electrodes.  The gate
can be included as a metallic plane at some distance $h$, and the
source-drain electrodes as a section of the wire with $\vF=0$. See
figure~2 for an illustration of this.
\begin{figure}
  \centerline{\includegraphics[width=0.5\textwidth]{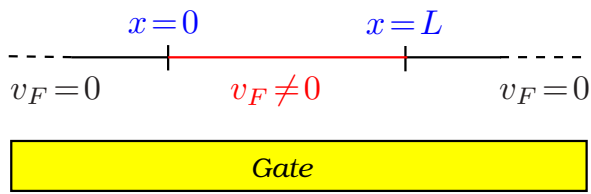}}
  \caption{The model used to find the electric. The electrodes are represented by a 1D lead with $v_F=0$.}
  \label{fig:model}
\end{figure}

To minimize the energy functional in \eqref{Frho} under the
constraint that a charge of one electron is added to the tube, we
add a Lagrange multiplier
\begin{equation}\label{Frholambda}
    F_1[\rho,\lambda]=E[\rho]+\lambda\left(\intL dx\rho(x)-e\right).
\end{equation}
First, we minimize with respect to $\rho$ and find
\begin{equation}\label{Erhomin}
    \frac{\delta F_1}{\delta \rho[x]}=
    \hbar\vF \rho(x)+\intL
    dx'\,V(x,x')\rho(x')+\lambda=0,
\end{equation}
and then with respect to $\lambda$:
\begin{equation}\label{Dlambda}
     \frac{\partial F_1}{\partial \lambda}=\intL dx\rho(x)-e=0.
\end{equation}
These two linear equations are readily solved numerically. Once the
solution is found, the electric field is given by
\begin{equation}\label{Exx}
     eE_x(x)=-\frac{\partial }{\partial x}\intL
    dx'\,V(x,x')\rho(x').
\end{equation}
The important parameters in this solution are the aspect ratio of
the tube, i.e., $\frac{L}{R}$, and the strength of the interaction
\begin{equation}\label{rs}
    r=\frac{e^2}{4\pi\epsilon_0 \hbar\vF}.
\end{equation}
For typical parameters, one has $r=10-20$, while the aspect ratio is
$200-2000$. The distance to the gate is not important, as long as it
longer than the length over which the electric decays, which is
typically the case. Our numerical solution gives an electric field
comparable to the results of Guinea\cite{guin05} (see also
reference~\cite{mish05}), which results  in an electric force  of
order $eE_x\sim 10^{-9}$ N, for typical nanotube devices. However,
this field is limited to a small region near the contacts, and
therefore the total effect of it is small.

For $r\gg 1$, we can in fact make a simple approximation which quit
accurately describes the full numerical solution. The electric field
can related to the charge density by differentiating the condition
(\ref{Erhomin}) with respect to $x$:
\begin{equation}\label{Exx2}
     eE_x(x)=\hbar\frac{\partial }{\partial x}\left[\rho(x)v_F(x)\right].
\end{equation}
Because $\rho(x)$ changes little along the wire, we may set
$\rho\approx 1/L$ and we thus obtain
\begin{equation}\label{Exx3}
     eE_x(x)\approx \frac{\hbar v_F }{L}\left[\delta(x)-\delta(x-L)\right].
\end{equation}
The width of the delta function will be given by microscopic details
of the interface between the tube and the contact, i.e. a length
scale of the order of the radius of the tube itself. This length
scale we denote by $x_0$.

\subsubsection{The electrostatic force in the longitudinal direction}

The term in the Hamiltonian describing the interaction between the
electron charge density and the nuclear system is
\begin{equation}\label{Helphlong}
    H_{\mathrm{el-vib},x}=-\int dxdx'\, \rho(x)V(x,x')\rho_n(x'),
\end{equation}
where  $\rho_n(x)$ is the density of the positive ions in the tube.
The force per length acting on the mechanical degrees of freedom is
therefore given by $eE_x\rho_n$. In terms of the displacement field
defined below in \eqref{udef}, $H_{\mathrm{el-vib},x}$ becomes
\begin{equation}\label{Helph1}
    H_{\mathrm{el-vib},x}=-\rho_0 \int dx\, \rho(x)V(x,x')\left[\partial_{x'} u(x')\right]=
    -e\rho_0 \int dx\,E_x(x)\, u(x).
\end{equation}

\subsection{Electric field perpendicular to the nanotube}
\label{sec:Eperp}

To find the electric field in the radial direction we model the
nanotube as a distributed capacitor similarly to Sapmaz et
al.\cite{sapm03}. We include capacitances to the electrodes $C_l$
and to the gate
\begin{equation}\label{Ctot}
    C=C_l+C_g,
\end{equation}
where the capacitance to the gate is
\begin{equation}\label{Cgate}
    C_g=\intL dx\, c(h(x)),\quad     c(h)=\frac{2\pi\epsilon_0}{\cosh^{-1}(h/R)},
\end{equation}
with $c$ being the distributed capacitance of a tube over a plane.

To find the total charge on the tube, we write the total
electrostatic energy as
\begin{equation}\label{Etotq}
    W=\frac{q^2}{2C}-q\Delta \Phi/e,
\end{equation}
where $C$ is total capacitance with charge $q$ and $\Delta \Phi$ is
the difference between the nanotube and the electrode work
functions. (Here we neglect the effect of the source, drain and gate
voltages, because they are considerably smaller than $\Delta \phi$.)
The optimum charge is thus
\begin{equation}\label{qopt}
    q_0\equiv n_0e=\Delta \Phi \,C/e.
\end{equation}
For single-walled carbon nanotubes the workfunction is about 4.7
eV\cite{gao01,zhao02,shan05}, while for gold it is 5.1 eV. For
typical devices $C\sim 10^{-17}$ F and hence  $n_0\sim 30$. The
electrostatic energy is used in the following section to calculate
force acting on the tube.

\subsubsection{The electrostatic force in the transverse direction}

Below we solve for the distortions of the wire due to the
electrostatic forces.  The force in the direction perpendicular to
the charged wire (denoted the $z$-direction) is given by
\begin{equation}\label{kedef}
    k=-\frac{d W}{dC_g}\left. \frac{\delta c_{g}}{\delta
z(x)}\right\vert_{z=0}=\,\left(\frac{C_g}{C}\right)^2\frac{e^2n_0^2}{4\pi\epsilon_0
hL^2},
\end{equation}
where
\begin{equation}\label{dCgdz}
\left.\frac{\delta c_{g}}{\delta z(x)}\right\vert_{z=0}=\frac{2\pi
\varepsilon
_{0}^{{}}}{h\left[\cosh^{-1}(h/R)\right]^{2}}=\frac{C_g^2}{2\pi\epsilon_0
hL^2}, \quad\mathrm{with}\,C_{g}=Lc(0).
\end{equation}

\section{Elastic properties}
\label{sec:elastic}

In this section, we discuss in detail the elastic properties of a
suspended nanotube. Most devices are made by under-etching after
depositum of the nanotube, which is done at room temperatures.
Therefore, since the thermal expansion coefficient of nanotubes is
negative (100 nm wire expands a few nm from room temperature to 4 K)
it seems unlikely that there is a large tension in tube unless it
becomes heavily charged\cite{sapm03}. When the tube is charged it is
attracted towards the metallic gates and leads.

The radial force, off course, couples to the breathing-type modes,
which, however, have too large energies ($\sim $ 30 meV) to be
relevant for the low-voltage experiment in reference~\cite{sapm05}.
Here we are interested in the lower part of the excitation spectrum
and disregard this mode. We are left with the bending and stretching
modes. The energy of the bending is typically much lower than those
of stretching modes\cite{ustu05}, and therefore we treat these as
purely classical, which means that we will solve for the bending
mode separately  and then via an anharmonic term this solution acts
as a force term on the longitudinal mode.

We thus consider two possible mechanism for coupling to the
stretching mode: either directly via the longitudinal electric field
discussed in section~\ref{sec:Epar} or through the perpendicular
field, section~\ref{sec:Eperp}, which bends the tubes and hence also
stretches it.

\subsection{Elasticity theory of a hanging string}

\begin{figure}
  \centerline{\includegraphics[width=0.5\textwidth]{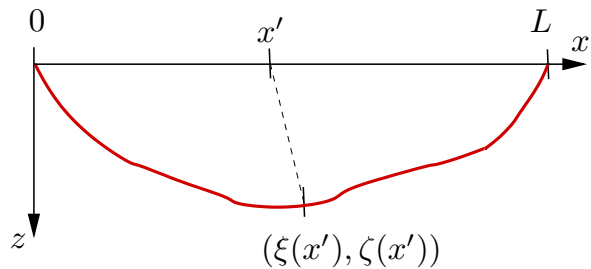}}
  \caption{The coordinate system used to describe the hanging tube. A point on the tube, which
before the distortion was at $(x',0)$ has after the deformation the
coordinates $(\xi(x'),z(x')).$}
  \label{fig:xizeta}
\end{figure}
Assuming the tube to lie in a single plane and thus having no
torsion in the longitudinal direction, we can describe the
distortion of the bend tube by $(\xi (x),z(x))$, where $x\in \lbrack
0,L]$ runs along the unbend tube in absence of external forces, see
figure~\ref{fig:xizeta}. (If the tube has some slack, $x$ is a curve
linear coordinate along the equilibrium position the hanging tube.)
This means that a point along the tube, which before was at $(x,0)$
after the deformation has the coordinates $(\xi (x),z(x)).$ The
total elastic energy of the tube is then follows from standard
elasticity theory of an elastic string\cite{landau:elasticity}
\begin{equation}\label{W}
W=\int_{0}^{L}dx\left(\frac{ EA[\zeta
(x)]^{2}}{2}+\frac{EI}{2[R(x)]^{2}}
+k_\perp^{{}}z(x)+k_\parallel^{{}}\;u(x)\right)
 ,
\end{equation}
where $\zeta $ is the linear strain of extension, $R$ is the radius
of curvature, $A=\pi R^2$ the area and $I=\pi R^4/4$ the moment of
inertia, $E$ is Young's modulus, and $k_\parallel,k_\perp$ the
external forces, and where we have defined the longitudinal
displacement field as
\begin{equation}\label{udef}
u(x)=\xi (x)-x.
\end{equation}
The linear extension of an infinitesimal element between $x$
and $x+dx$ is
\begin{equation}
\zeta (x)dx=\sqrt{(\xi (x+dx)-\xi (x))^{2}+(z(x+dx)-z(x))^{2}}-dx,
\end{equation}
or
\begin{equation}
\zeta (x)=\left( \sqrt{[1+u^{\prime }(x)]^{2}+[z^{\prime
}(x)]^{2}}-1\right).
\end{equation}
The linear extension elastic energy is thus
\begin{equation}\label{Wlin}
W_{\mathrm{lin}}=\frac{EA}{2}\int_{0}^{L}dx\left( \sqrt{[1+u^{\prime
}(x)]^{2}+[z^{\prime }(x)]^{2}}-1\right)^2 .
\end{equation}

The curvature contribution is determined in a similar way. First,
the unit tangential vector is
\begin{equation}
\mathbf{t=}\frac{\left( \xi ^{\prime }(x),z^{\prime
}(x)\right)}{\sqrt{[\xi ^{\prime }(x)]^{2}+[z^{\prime }(x)]^{2}}} ,
\end{equation}
which then gives the square of the radius of curvature as
\begin{equation}
R^{-2}=\left(\frac{d\mathbf{t}}{dl}\right)^{2}=\left(\frac{d\mathbf{t}}{dx}
\frac{dx}{dl}\right)^{2},\quad \frac{dl}{dx}=\sqrt{[1+u^{\prime
}(x)]^{2}+[z^{\prime }(x)]^{2}},
\end{equation}
and then the curvature contribution to the elastic energy finally
becomes
\begin{equation}\label{Wcurv}
    W_{\mathrm{curv}}=\frac{EI}{2}\int_{0}^{L}dx\,
    \frac{(z'(x)u''(x)-(1+u'(x))z''(x))^2}{([1+u'(x)]^2+[z'(x)]^2)^3}.
\end{equation}

\subsection{Weak distortions}

Since we are interested in small deflections, we expand the two
elastic energy expressions for small $z$ and $u.$ For
$W_{\mathrm{lin}}$, we obtain to third order in $u $ and $z$
\begin{equation}\label{Wlinf}
W_{\mathrm{lin}}\approx \frac{EA}{2}\intL\,dx\left( [u^{\prime
}(x)]^{2}+\frac{[z^{\prime }(x)]^{4}}{4}+u^{\prime }(x)[z^{\prime
}(x)]^{2}\right) .
\end{equation}
Here we note that the last term couples the bending and stretching
modes. For the curvature contribution, we find to the same order
\begin{equation}
W_{\mathrm{cur}}=\frac{EI}{2}\intL dx\,\left( 1-4u^{\prime
}(x)\right) \left[ z^{\prime \prime }(x)\right] ^{2}\approx
\frac{EI}{2}\intL dx\,\left[ z^{\prime \prime }(x)\right] ^{2}.
\end{equation}
Again, there is a term which couples the two modes. However, for
nanotubes this term is  much smaller than the last term in
\eqref{Wlinf}, because it smaller by a factor $(R/L)^2$, and hence
we have neglected the coupling term in $W_\mathrm{cur}$.

\section{Solution for the bending of the tube}

As mentioned above, the bending mode itself has a resonance
frequency too low to be seen tunnelling spectroscopy ($\sim 100$
MHz\cite{ustu05})(even when under tension due to the charging of the
wire ), which means that it can be treated as a classical degree of
freedom and  Franck-Condon type physics is not involved. In
contrast, the longitudinal stretching mode has been seen in the
single-electron-transistor (SET) device\cite{sapm05} and here we
wish to calculate the Franck-Condon coupling constants for this
mode. Therefore we take the following approach: first we solve for
the bending mode classically and then insert this as an external
force acting on the longitudinal mode. The differential equation for
$z(x)$ is
\begin{equation}
IEz^{\prime \prime \prime \prime }-\frac{AE}{2}(z^{\prime
})^{2}z^{\prime \prime }=k.  \label{eqom}
\end{equation}
This equation cannot be solved analytically. One approach is to
approximate $(z^{\prime })^{2}$ by the average value, which is
corresponds to assuming constant tension in the wire\cite{tension}.

Below we solve for the bending function $z(x)$ in two regimes: the
linear and the non-linear regime. Once we know $z(x)$, we will be
interested in the \textit{change of $z(x)$ due to tunnelling of a
single electron}. For large $n_0$, the relevant change is thus
\begin{equation}\label{zegendef}
    z_e(x)=\frac{dz(x)}{dn_0}.
\end{equation}
This change will then couple to the longitudinal mode via the
coupling term in \eqref{Wlinf}.

\subsection{Linear regime}

For weak forces we can simply neglect the non-linear term in
\eqref{eqom}, and with boundary conditions $z(0)=z(L)=z^{\prime
}(0)=z^{\prime }(L)=0$ the solution is
\begin{equation}\label{z0def}
z_0(x)=\frac{kL^{4}}{24EI}\left( 1-\frac{x}{L}\right) ^{2}\left(
\frac{x}{L}\right)^{2}.  \label{smallK}
\end{equation}

The shift in $z_0(x)$ due to the charge of a single electron is
according to \eqref{zegendef} given by
\begin{equation}\label{zedef}
z_{0,e}(x)=\frac{e^2 n_0}{12\pi\epsilon_0 h}\frac{1}{E
A}\left(\frac{L}{R}\right)^2\left(\frac{C_g}{C}\right)^2\left(
1-\frac{x}{L}\right) ^{2}\left( \frac{x}{L}\right)^{2}.
\label{smallKe}
\end{equation}
For a tube with $R=$ 0.6 nm, $L=1\,\mu$m, and $E \approx 10^{12}$ Pa
and a device with $h=200$ nm, $n_0=50$, and $C_g/C=0.5$, the maximum
distortion is of order a few nm.

The linear approximation is valid when the second term in
\eqref{eqom} is much smaller than the first. Using the solution in
\eqref{smallK}, this translates to the condition
\begin{equation}\label{condition}
    10^{-6}\left(\frac{L}{R}\right)^2\left(\frac{kL^3}{EI}\right)^2\ll 1.
\end{equation}
For the typical parameters used below, the number on the left hand
side of (\ref{condition}) is $\lesssim 1$, and therefore the linear
approximation is only marginally valid.

\subsection{Non-linear regime.}
For larger distortions the non-linear term in \eqref{eqom} becomes
important. In the strongly non-linear regime, we can neglect the
first term and we have
\begin{equation}\label{eqomnonlin}
\frac{AE}{2}(z^{\prime })^{2}z^{\prime
\prime}=\frac{AE}{6}\frac{d}{dx}(z^{\prime })^{3}=-k,
\end{equation}
with boundary condition  $z(0)=z(L)=0$. The solution of this
equation is
\begin{equation}\label{znonlin}
    z_1'(x)= \left(\frac{6kL^3}{EA}\right)^{1/3}\left|\frac{x}{L}-\frac12\right|^{1/3}
    \mathrm{sign}\left(\frac{L}{2}-x\right).
\end{equation}
In this non-linear regime, the change in the slope the bending
function $z_1(x)$ due to a single electron charge, is
\begin{equation}\label{ze1def}
    z_{1,e}'(x)=2\left(\frac{C_g}{C}\right)^{2/3}\left(\frac{e^2}{2\pi\epsilon_0
hAn_0}\right)^{1/3}\left|\frac{x}{L}-\frac12\right|^{1/3}
    \mathrm{sign}\left(\frac{L}{2}-x\right).
\end{equation}

\section{Distortion of the longitudinal mode}

\begin{figure}
  \centerline{\includegraphics[width=0.4\textwidth]{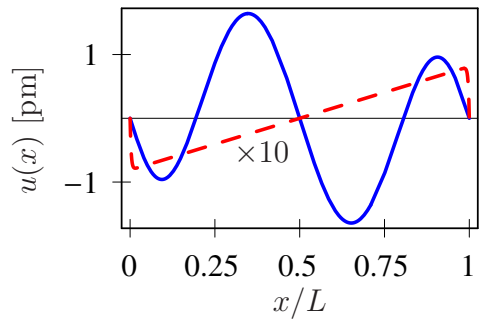}}
  \caption{The solutions for the shifted longitudinal mode $u(x)$ due to the parallel (red dashed curve) and
  perpendicular (blue curve) electric fields. We have used typical parameters as in section~\ref{sec:fc}}
  \label{fig:u}
\end{figure}

Due to the electrostatic forces the equilibrium position of
longitudinal displacement field $u(x)$ shifts. Since the forces are
small the displacement is going to be small. For the tunnelling
overlap factors, the important point is, however, whether these
displacements are large compared to the quantum uncertainty length,
which later is seen to be of order of pm. In this section, we
calculate the classical displacements of $u(x)$. One example is
shown in figure~\ref{fig:u}.

\subsection{Distortion due to the longitudinal electrostatic force}

The displacement of the longitudinal mode is readily found from its
equation of motion. The displacement due to the longitudinal
electric field follows from
\begin{equation}\label{ulong}
EA u''_0(x)=k_\parallel(x)=-eE_x\rho_0,
\end{equation}
with boundary conditions $u(0)=u(L)=0$. With forces concentrated
near the contacts as in \eqref{Exx3}, there is an abrupt change of
$u(x)$ at $x=0$ and $x=L$.  See figure~\ref{fig:model} red dashed
curve.

\subsection{Distortion due to the transverse electrostatic force}

Once we have solved for $z(x)$ in \eqref{zegendef}, we can find the
force that acts on the longitudinal diplacement field $u$ by
inserting the solution into the last term of \eqref{Wlinf}, and then
identify the force $k_\perp$ in \eqref{W}. This gives
\begin{equation}\label{Kudef}
    k_\perp=\frac{d}{dx}\frac{EA}{2}[z'(x)]^2.
\end{equation}
The size of the displacement follows from the balance between this
force and the strain:
\begin{equation}\label{usolve}
EA u''_0(x)=k_\perp(x) \Leftrightarrow  u''_0(x)=z'(x)z''(x),
\end{equation}
which together with the boundary condition, $u_0(0)=u_0(L)=0$, gives
the solution
\begin{equation}\label{usolvef}
u_0(x)=-\frac12\int_0^x dy \,[z'(y)]^2+\frac{x}{2L}\intL
dy\,[z'(y)]^2.
\end{equation}
One example is shown in figure~\ref{fig:model} (blue curve).

In the next section, we analyze the Franck-Condon overlap factor due
to this electrostatic distortion of the stretching mode.

\section{Quantum mechanics and Franck-Condon overlap factors}

\label{sec:FC}

The longitudinal eigenmodes of a nanotube modeled as a 1D elastic
medium follows from the Hamiltonian
\begin{equation}
\hat{H}=\int_{0}^{L}dx\left( \frac{\hat{p}^{2}(x)}{2\rho
_{m}^{{}}}+\frac{AE }{2}(\partial _{x}\hat{u}(x))^{2}\right) ,
\end{equation}
where $\rho _{m}$ is mass density per unit length, and $\hat{u}$ and
$\hat{p}$ are conjugated variables, i.e.,
$[\hat{u}(x),\hat{p}(x')]=i\hbar\delta(x-x')$. In order to
diagonalize $\hat{H}$, we introduce the Fourier transforms:
\begin{eqnarray}
\hat{u}(x) &=&\sqrt{2}\sum_{n=1}^{\infty }\sin \left( \frac{\pi
nx}{L} \right) \hat{u}_{n}^{{}},\quad \hat{u}_{n}=\frac{\sqrt{2}}{L}
\int_{0}^{L}dz\sin \left( \frac{\pi nx}{L}\right) \hat{u}(x), \\
\hat{p}(x) &=&\frac{\sqrt{2}}{L}\sum_{n=1}^{\infty }\sin \left(
\frac{\pi nx }{L}\right) \hat{p}_{n}^{{}},\quad
\hat{p}_{n}=\sqrt{2}\int_{0}^{L}dz\sin \left( \frac{\pi
nx}{L}\right) \hat{p}(x),
\end{eqnarray}
where $\hat{u}_{n}$ and $\hat{p}_{n}$ obey
$[\hat{u}_{n},\hat{p}_{n}]=i\hbar$. Now $\hat{H}$ transforms to
\begin{equation}\label{Hunpn}
\hat{H}=\sum_{n=1}^{\infty }\left(
\frac{\hat{p}_{n}^{2}}{2M}+\frac{AE}{2L} \left( \pi n\right)
^{2}\hat{u}_{n}^{2}\right) ,\quad M=\rho _{m}^{{}}L,
\end{equation}

The Hamiltonian (\ref{Hunpn}) is easily diagonalized by
\begin{equation}
\hat{u}_{n}^{{}}=\ell _{0,n}\sqrt{\frac{1}{2}}\left(
\hat{a}_{n}^{{}}+\hat{a} _{n}^{\dagger }\right) ,\quad
\hat{p}_{n}^{{}}=\frac{i}{\ell _{0,n}}\sqrt{ \frac{1}{2}}\left(
\hat{a}_{n}^{\dagger }-\hat{a}_{n}^{{}}\right) ,
\end{equation}
where $\hat{a}_n^{{}}$ and $\hat{a}_n^\dagger$ are usual
annihilation and creation operators, and
\begin{equation}
\Omega =\pi \sqrt{\frac{AE}{ML}}=\frac{\pi }{L}\sqrt{\frac{AE}{\rho
_{m}^{{}} }}\equiv \frac{\pi v_s}{L},\quad \ell
_{0,n}=\sqrt{\frac{\hbar }{nM\Omega }}.
\end{equation}
With these operators, the Hamiltonian (\ref{Hunpn}) becomes
\begin{equation}
\hat{H}=\sum_{n=1}^{\infty }\hbar \Omega n\left(
\hat{a}_{n}^{\dagger }\hat{a}_{n}^{{}}+\frac12\right) .
\end{equation}

As we saw in the previous sections, additional terms in the
Hamiltonian appears due to the force generated by the tunnelling
electron. These are included next.

\subsection{Coupling due to longitudinal electric field}

The longitudinal electric field $E_x$ gives rise to a coupling
Hamiltonian (see \eqref{Helph1}):
\begin{equation}\label{Hparal}
\hat{H}_{\mathrm{el-vib},\parallel}=-e\rho_0\int_{0}^{L}dx~\hat{u}(x)E_x(x)
=\sum_{n=1}^\infty \hat{u}_n f_{n,\parallel},
\end{equation}
where
\begin{equation}\label{fparal}
    f_{n,\parallel}^{{}}= -e\rho_0\sqrt{2}\int_{0}^{L}dx~\sin
\left( \frac{\pi nx}{L} \right) E_x(x)\approx-(2\pi\sqrt{2})\frac{
n\hbar\vF\rho_0x_0}{L^2},
\end{equation}
for $n$ even and zero for $n$ odd.

\subsection{Coupling due to the capacitative force}

The transverse force leads to the following term in the Hamiltonian
(see \eqref{Wlinf}):
\begin{equation}\label{Hperp}
\hat{H}_{\mathrm{el-vib},\perp}=\frac{EA}{2}\int_{0}^{L}dx~\hat{u}^{\prime
}(x)[z^{\prime }_e(x)]^{2}=\sum_{n=1}^{\infty
}\hat{u}_{n}^{{}}f_{n,\perp}^{{}}.
\end{equation}
where
\begin{equation}\label{fnperp}
f_{n,\perp}^{{}}= \frac{EA}{\sqrt{2}}\frac{n\pi
}{L}\int_{0}^{L}dx~\cos \left( \frac{\pi nx}{L} \right) [z^{\prime
}_e(x)]^{2}.
\end{equation}

\subsection{Franck-Condon overlap factors}
\label{sec:fc}

The tunnelling of an electron leads to a displacement of the
equilibrium displacement field according to
equations~(\ref{Hparal}),(\ref{Hperp}). Each mode represented by
$\hat{u}_n$ is shifted by the amounts
\begin{equation}\label{elln}
\ell _{n,a}= \frac{Lf_{n,a}^{{}}}{AE(\pi n)^{2}},\quad
a=(\parallel,\perp).
\end{equation}
This allows us to calculate the Franck-Condon parameters, which for
each mode $n$ expresses  the overlaps between the new eigenstates
around the new equilibrium positions and the old one. This parameter
is defined as\cite{brai03flen}
\begin{equation}\label{gndefa}
    g_{n,a}=\frac{1}{2}\left(\frac{\ell_{n,a}}{\ell_{0,n}}\right)^2.
\end{equation}
The size of $g$ determines the character of the vibron sidebands in
the $IV$-characteristics, such that for $g\ll 1$ there are no
sidebands, for $g$ of order one clear sidebands are seen, while for
$g\gg 1$ a gap appears in $IV$
characteristic\cite{brai03flen,koch05}.

For the Franck-Condon parameter due to the parallel electric field,
we thus have
\begin{equation}\label{gndefp}
    g_{n,\parallel}=\frac{4}{\pi^2n^2}\,\frac{M\Omega}{\hbar}\left(\frac{\hbar\vF\rho_0x_0}{AEL}\right)^2.
\end{equation}
Using $x_0\approx 1$ nm, $L=500$ nm, $\vF=10^6$ ms$^{-1}$, $R=0.6$
nm, and $E=10^{12}$ Pa, we find $g_{2,\parallel}\sim 10^{-5}$. This
is clearly too small a coupling constant to explain the experimental
findings in reference~\cite{sapm05}.

The coupling constant due to the perpendicular electric can be
expressed explicitly, for the case of small $z(x)$, using
\eqref{z0def}. The integral in \eqref{fnperp} can then be performed
and we obtain
\begin{equation}\label{gperp}
g_{n,\perp}=\frac{M\Omega L^{2}}{\hbar}\left( \frac{k_eL^{3}}{EI}
\right) ^{4}\frac{\left(n^{2}\pi^{2}-40\right)^{2}}{8 n^{13}\pi
^{14}},\quad \mathrm{for}\quad n\quad \mathrm{even},
\end{equation}
and zero for $n$ odd. Using $k_e$ as defined in \eqref{kedef}, and
typical parameter for single wall carbon nanotube devices: $R=0.6$
nm, $L=1\,\mu$m, $E=10^{12}$ Pa, $h\simeq 10-200$ nm, $n_0=L
c(0)\Delta\phi\sim 30$, $C_g/C=0.1-0.75$, we find the maximum $g_n$
factor to occur for $n=4$. However, for this range of parameters we
also find $g_4\ll 1$, unless the geometry is such that
\begin{equation}\label{geo}
\frac{n_0\alpha^2}{h [\nm]}>0.1.
\end{equation}
Even though we can get a significant coupling, the condition
(\ref{geo}) does not seem to be compatible with the experimental
realizations in reference~\cite{sapm05}. Even more so because the
coupling strongly strength, \eqref{gperp}, depends strongly on the
length of wire, which is not seen experimentally.

\section{Conclusion and discussion}
\label{sec:disc}

We have considered the electromechanics of suspended nanotube
single-electron-transistor devices. When the charge on the tube is
changed by one electron the resulting electric field will distort
the tube in both the longitudinal and transverse directions, and
both these distortions couple to the stretching mode. We have
calculated they consequences for the coupling constant for
vibron-assisted tunnelling expressed as the Franck-Condon factor.
This is expressed in terms of the ratio of the classical
displacement to the quantum uncertainty length. Even though both are
in the range of picometers, the effective coupling parameters, $g$,
turn out to be small for most devices.

Because the screening of the longitudinal electric field is very
effective, the dominant interaction seems to be the coupling via the
bending mode. However, only if the tube is very close to the gate do
we get a sizeable $g$-parameter. This could indicate that in the
experiment of Sapmaz et al.\cite{sapm05} the suspended nanotube has
a considerable slack, which would diminish the distance to the gate.
Further experiments and more precise modelling of actual geometries
should be able to resolve these issues.

\ack \vspace{-.25cm} We thank the authors of reference~\cite{sapm05}
for valuable discussions. The work is supported in part by the
European Commission through project FP6-003673 CANEL of the IST
Priority.
\appendix
\vspace{-.25cm}

\section{Electric field in a charged nanowire}
\label{app:electro} \vspace{-.25cm}
\subsection{The effective 1D interaction}

Here we derive in more detail the result show in
section~\ref{sec:electrostat}. The interaction between two charges
on points $x$ and $x'$ is
\begin{equation}\label{Vxx}
    V(x-x')=\int d\mathbf{r}_\perp\int
    d\mathbf{r}_\perp'\frac{e^2|\phi({r}_\perp)|^2|\phi({r}_\perp')|^2}
    {4\pi\epsilon_0\sqrt{(x-x')^2+(\mathbf{r}_\perp-\mathbf{r}_\perp')^2}},
\end{equation}
where $\phi(\mathbf{r}_\perp)$ is the wavefunction is the
perpendicular direction. By modelling the tube as a cylinder,
$\phi(\mathbf{r}_\perp)=\delta(r-R)/2\pi R$, we get
\begin{eqnarray}\label{Vxx2}
    V(x-x')&=&\frac{e^2}{2\pi\epsilon_0}\int_0^{2\pi}
    \frac{d\theta}
    {\sqrt{(x-x')^2+(2R\sin(\theta/2)^2}}\nonumber\\
    &=&\frac{2e^2}{2\pi^2\epsilon_0\sqrt{(x-x)^2+4R^2}}\,K\!\left(\frac{4R^2}{(x-x')^2+4R^2}\right),
\end{eqnarray}
where $K$ is the complete elliptic integral of the first kind. With
screening due to a gate at distance $h\gg R$, the interaction is
\begin{equation}\label{Vsc}
    V_\mathrm{scr}(x-x')=V(x-x')-\frac{e^2}
    {4\pi\epsilon_0\sqrt{(x-x')^2+(2h)^2}}.
\end{equation}


\begin{thebibliography}{10}
\bibitem{tension}
The constant tension approximation is ony valid for forces
concentrated to points\cite{landau:elasticity}, which is not the
case here where the force is distributed. This is in constrast to
the assumptions made in \cite{sapm03}.
\end{thebibliography}
\end{document}